\documentstyle[twocolumn,prb,aps]{revtex}
\begin{document}
\title{Localization in Semiconductor Quantum Wire Nanostructures}
\author{Dongzi Liu\cite{addr} and S. Das Sarma}
\address{Department of Physics,
University of Maryland,
College Park, Maryland 20742-4111}
\address{\rm (Submitted to Physical Review B  on 23 December 1994)}
\address{\mbox{ }}
\address{\mbox{ }}
\address{\parbox{14cm}{\rm \mbox{ }\mbox{ }\mbox{ }
Localization properties of quasi-one dimensional
quantum wire nanostructures are investigated using the transfer
matrix-Lyapunov exponent technique.
We calculate the  localization length as a function of the effective
mean-field mobility assuming the random disorder potential to be arising
from dopant-induced
short-range $\delta$-function or finite-range Gaussian impurity
scattering.
The localization length increases approximately
linearly with the effective mobility, and is also enhanced by
finite-range disorder. There is a sharp reduction in the localization
length when the chemical potential crosses into the second subband.
}}
\address{\mbox{ }}
\address{\mbox{ }}
\address{\parbox{14cm}{\rm PACS numbers: 73.20.Dx; 71.30.h; 73.20.Jc.}}
\maketitle

\makeatletter
\global\@specialpagefalse
\def\@oddhead{REV\TeX{} 3.0\hfill Das Sarma Group Preprint, 1994}
\let\@evenhead\@oddhead
\makeatother

An early \cite{sakaki} motivation in developing semiconductor quantum
wires was the suggestion that one-dimensional phase space restrictions
would severely reduce impurity scattering, thereby substantially
enhancing electron mobilities making possible faster transistors and
optoelectronic devices. The rationale for this proposed mobility
enhancement is the fact that at low temperatures the only possible
resistive scattering process in one dimension is the
$2k_F$-scattering, implying enhanced low temperature mobility as all
other scattering processes are suppressed. In real quasi-one
dimensional quantum wire systems there would be additional
inter-subband scattering processes which are, however, weak in
general, and therefore, it was argued \cite{sakaki} that high quality
quasi-one dimensional GaAs quantum wires could potentially have low
temperature mobilities surpassing those in modulation-doped two
dimensional HEMT structures. There is, however, a serious flaw in this
phase-space-restriction-induced mobility enhancement argument:
Impurity disorder-induced $2k_F$ multiple scattering leads, in fact,
to localization and zero mobility in one dimensional systems. It is
well-established that all one electron states in a disordered one
dimensional system are exponentially localized \cite{ishii}, and this
Anderson localization phenomenon \cite{anderson} leads to the
inevitable rigorous conclusion that strictly one dimensional quantum
wire structures are always zero mobility insulators (because some
disorder must always be present in real systems), and cannot carry any
current in the thermodynamic limit. In high-quality low-disorder
quantum wires, the localization length may be large, and only if the
length of the wire is shorter than the localization length, the system
can behave as an ``effective'' metal with non-zero effective
mobilities.
The important issue for quantum wire transport is then to figure out
the localization length for a given realization of disorder, which
could be parameterized by the mean-field mobility calculated within the
Born approximation which being a single impurity scattering
approximation
does not lead to localization (localization being \cite{anderson} a purely
multiple
scattering induced interference phenomenon).

In this paper, we carry out a calculation of the quantum wire
localization length as a function of disorder using the mean-field
Born approximation mobility as the relevant parameter characterizing
the impurity scattering strength. We find that in reasonable
high-quality GaAs quantum wires the localization length may be several
microns, making it possible to fabricate current carrying quantum
wires for nanoelectronic or even microelectronic applications.

Our goal is to calculate the zero temperature quantum wire
localization length in the presence of random impurity disorder.
We model the impurities as randomly distributed point scattering
centers with the electron-impurity interaction
as short-range $\delta$-function or finite-range Gaussian potentials.
The mean field
mobility of the quantum wire is calculated using the first order Born
approximation, which should be roughly the same as the corresponding
two dimensional mobility of the GaAs material from which the quantum
wire is made.
 The electronic localization length at the Fermi
level is determined by calculating the Lyapunov exponent of the transfer
matrix for the system \cite{lyapunov}.
 We find that the localization length increases
linearly with the mobility of the wire at low electron density ({\it
i.e.} when only the lowest subband is occupied).
Other things being equal ({\it i.e.} the same mobility)
finite-range of the scattering potential
 tends to increase the localization length. We also find
a sharp reduction in the localization length
when the electrons start to fill the
second subband.


The quantum wire is modeled as a two-dimensional strip along the $x$ direction
with a length $L$ and a width $w$ along the $y$ direction
with randomly distributed scattering centers along the strip.
Without any loss of generality, we
ignore the third direction ({\it i.e.} the $z$ axis) completely
because the thickness of real GaAs quantum wires is often much less
than their width. For a perfect quantum wire with an infinite
rectangular confining potential well along the $y$ direction, the
energy spectrum and the electron wavefunction can be written as:
\begin{mathletters}
\begin{eqnarray}
E_{nk} & = &
\frac{\hbar^2k^2}{2m^*}+\frac{\hbar^2}{2m^*}(\frac{n\pi}{w})^2, \\
\phi_{nk}& =
&\frac{1}{\sqrt{L}}e^{ikx}\sqrt{\frac{2}{w}}\sin{(\frac{n\pi}{w}y)}.
\end{eqnarray}
\end{mathletters}
The impurity potential centered at $(x_i,y_i)$ is taken to be:
\begin{mathletters}
\begin{equation}
 V_{im}(x,y)= v_o\delta(x-x_i)\delta(y-y_i)\label{imp-a}
\end{equation}
or
\begin{equation}
 V_{im}(x,y) = \frac{v_o}{\pi
s^2}e^{-(x-x_i)^2/s^2}e^{-(y-y_i)^2/s^2}\label{imp-b}.
\end{equation}
\end{mathletters}
The infinite potential well confinement is obviously an idealization
which, for our purpose, should be a good approximation \cite{lai-ds} because
choosing a different model confinement will only change the impurity
potential matrix elements {\it i.e.} the disorder strength $v_o$ which
we are parameterizing. While our choice for the impurity potential in
Eqs.(\ref{imp-a}) and (\ref{imp-b}) is mainly a matter of convenience,
we have explicitly verified \cite{ds-hu-liu} that screened Coulomb
impurity potential in a GaAs quantum wire may be reasonably
approximated by a Gaussian potential. We also make a one- or two-subband
approximation for our calculations which should be accurate for
low electron densities. Including more (than two) subbands should not
change our results qualitatively.

The localization length is calculated using the transfer matrix
technique \cite{lyapunov}. We first divide
 the wire of length $L$ into small segments of length $a$ (which is
taken to be  much
less than the average inter-impurity distance). The system can then be
effectively described by the two-channel (corresponding to two
subbands)
Anderson model with nearest
neighbor hopping, {\it i.e.}
\begin{eqnarray}
\cal{H}&=&\sum_i\sum_{n,m=1,2}(E_n\delta_{nm}+V_{nm})|ni><mi| \\
 & &\mbox{ }-
\sum_i\sum_{n=1,2} t(|ni><ni+1|+|ni+1><ni|)\nonumber
\end{eqnarray}
where the hopping term $t=\hbar^2/2m^*a^2$, and $V_{nm}$ are the
random on-site
scattering matrix elements which can be calculated from the
impurity potential and the electron wavefunction.
A transfer matrix formalism \cite{lyapunov} can be easily set
up for such a Hamiltonian. The localization length is then given by the
inverse of the smallest Lyapunov exponent of the transfer matrix
\cite{lyapunov}.

The mean field mobility of the wire, which is used to parameterize the
strength of disorder, is calculated using the first order Born
approximation. If only the lowest subband is occupied, an electron can
only be scattered from $k_F$ to $-k_F$ or vice versa. The transport scattering
rate is then  given by:
\begin{equation}
\frac{1}{\tau}=\frac{2m^*N_i}{\hbar^3k_F}<|V_{11}(2k_F)|^2>,
\label{t1}
\end{equation}
with the mobility $\mu=e\tau/m^*$, where the matrix element $V_{11}$
of the impurity potential $V_{im}$ is taken entirely within the
lowest subband $n=1$. The mobility calculation for the 2-subband
scattering problem is discussed below. Because of the exact one to one
correspondence between the random disorder and the mean-field
mobility, $\mu\propto N_i<|V|^2>$, as defined in Eq.(\ref{t1}), the
impurity potential $V_{im}$ may be uniquely parameterized by the mean
field mobility $\mu$ provided the average impurity density $N_i$ is
known. We take $N_i$ to be equal to the average electron density
(again with no loss of generality) in the quantum wire, which is taken
to be $10^6cm^{-1}$ for all our calculations.


\begin{figure}
 \vbox to 5.0cm {\vss\hbox to 6cm
 {\hss\
   {\includegraphics{/home/ldz/tek/paper/psd05/f1.ps}
   }
  \hss}
 }
\caption{
The relation among the localization length($\lambda_L$),
mobility($\mu$) and the range of the impurity potential($s$) for a
$500\mu m$ long and $100\AA $ wide wire. The impurity concentration is
$1\times 10^6cm^{-1}$ (the same as the electron density).
\label{wire-f1}
}
\end{figure}

Our one-subband localization calculation results in the lowest subband are
shown
in Fig.\ref{wire-f1}. The short-range scattering model corresponds to
the $s=0$ limit. The mean field mobility in Fig.\ref{wire-f1} is calculated
neglecting inter-subband scattering, which we discuss below.
As shown in Fig.\ref{wire-f1},
the localization length increases approximately
linearly with the mobility of the wire. Finite-range scatterers
also tend to increase the localization length.
It is well known that the Lyapunov exponent takes a long time to
converge in the Anderson model, particularly for finite range
disorder.
The results shown in Fig.\ref{wire-f1} involve upto $10^7$ iterations.
It is noteworthy that
we still have fluctuations in our results for high mobility (and
finite potential range) samples even after $10^7$ iterations!
In general, the localization length is found to be larger than the
elastic mean free path extracted from the mean-field mobility.

As can be seen from  Eq.(\ref{t1}), the electron mobility can be increased by
increasing the electron density which is proportional to $k_F$
(assuming, of course, that $N_i$ is fixed). This
is true when electrons only occupy the lowest subband. As
electrons start to occupy the second subband, enhanced inter-subband
scattering is
introduced at the Fermi level \cite{ds-xie} which could result
in a sharp reduction of the localization length.
For example, an electron with momentum $k_{F1}$ in the lowest subband could be
scattered into either the $-k_{F1}$ state in the same subband or the $\pm
k_{F2}$
states in the second subband.   The two-subband generalization of
Eq.(\ref{t1}) is straightforward.
The scattering time for the first subband $\tau_1$ is different from
that for the second subband $\tau_2$. Using the
Boltzmann equation and first order Born approximation,
it is straightforward to show that in the most general two-subband
model including inter-subband scattering processes:
\begin{mathletters}
\begin{eqnarray}
\frac{1}{\tau_1}&=&\frac{2m^*N_i}{\hbar^3k_{F1}}<|V_{11}(2k_{F1})|^2>\\
& &\mbox{ }
+\frac{m^*N_i}{\hbar^3k_{F2}}\left[(1-\frac{k_{F2}\tau_2}{k_{F1}\tau_1})<|V_{12}(k_{F1}-k_{F2})|^2>\right.\nonumber\\
& &\mbox{ }\mbox{ }
+\left.
(1+\frac{k_{F2}\tau_2}{k_{F1}\tau_1})<|V_{12}(k_{F1}+k_{F2})|^2>\right],
\nonumber\\
\frac{1}{\tau_2}&=&\frac{2m^*N_i}{\hbar^3k_{F2}}<|V_{22}(2k_{F2})|^2>\\
& &\mbox{ }
+\frac{m^*N_i}{\hbar^3k_{F1}}\left[(1-\frac{k_{F1}\tau_1}{k_{F2}\tau_2})<|V_{21}(k_{F1}-k_{F2})|^2>\right.\nonumber\\
& &\mbox{ }\mbox{ }
+\left.
(1+\frac{k_{F1}\tau_1}{k_{F2}\tau_2})<|V_{21}(k_{F1}+k_{F2})|^2>\right].
\nonumber
\end{eqnarray}
\end{mathletters}
\noindent The mobility is then easily calculated once we get $\tau_1$ and
$\tau_2$.
It is interesting to note that when the second subband just starts to be
occupied({\it i.e.} $k_{F2}$ is very small), both $\tau_1$ and
$\tau_2$ are small leading to a mobility reduction due to enhanced
inter-subband scattering \cite{ds-xie}.
This causes a sharp decrease in the localization length as $E_F$
enters the second subband.

We present our two-subband localization results
 in Fig.\ref{wire-f2} for a $200\AA$ wide quantum wire.
The impurity concentration is fixed at $10^6cm^{-1}$. We use a fixed impurity
strength $v_o$ which is normalized
to the results presented in Fig.\ref{wire-f1} with $\delta$-function
impurity and a mobility value of $1\times 10^5cm^2V^{-1}s^{-1}$. Both
the mobility
and the localization length drop sharply
when electrons start to fill the second subband.
It is interesting to note that (similar to Fig.\ref{wire-f1}) the
localization length is almost an order of magnitude larger than the
effective mean free path obtained from the Born approximation mobility
values.

\begin{figure}
 \vbox to 5.0cm {\vss\hbox to 6cm
 {\hss\
   {\includegraphics{/home/ldz/tek/paper/psd05/f2a.ps}
   }
  \hss}
 }
\vspace{1mm}
 \vbox to 5.0cm {\vss\hbox to 6cm
 {\hss\
   {\includegraphics{/home/ldz/tek/paper/psd05/f2b.ps}
   }
  \hss}
 }
\vspace{1mm}
 \vbox to 5.0cm {\vss\hbox to 6cm
 {\hss\
   {\includegraphics{/home/ldz/tek/paper/psd05/f2c.ps}
   }
  \hss}
 }
\vspace{1mm}
 \vbox to 5.0cm {\vss\hbox to 6cm
 {\hss\
   {\includegraphics{/home/ldz/tek/paper/psd05/f2d.ps}
   }
  \hss}
 }
\caption{
The variation of density of states(a), electron mean free path(b),
mobility(c), and localization length(d) as the Fermi energy (measured
in units of $\Delta$) increases.
($\Delta$ is the energy difference between the bottom of the lowest two
subbands.)
The wire is $500\mu m$ long and $200\AA $ wide.
Results for three different impurity potential ranges
 are shown: $0$(solid line),
$2nm$(dashed line), and $5nm$(dotted line).
The impurity concentration is
$1\times 10^6cm^{-1}$. The impurity potential strength is normalized
to that of the wire presented in Fig.\protect\ref{wire-f1} with
$\delta$-function
impurity and $1\times 10^5cm^2V^{-1}s^{-1}$ mobility.
\label{wire-f2}
}
\end{figure}

In summary, we have calculated the localization length  for
a quasi-one dimensional quantum wire within the two-subband model.
We find that the localization length increases
linearly with the mobility of the wire at low electron density ({\it
i.e.} only the lowest subband is occupied). Finite-range scatterers
also tend to increase the localization length.
The sharp drop in the localization length seen in Fig.\ref{wire-f2}
may be difficult to observe experimentally because of finite thermal
and collisional broadening effects which may affect the one
dimensional density of states singularities of Fig.\ref{wire-f2}(a).
The localization length, in general, is found to be much larger than
the elastic mean free path calculated on the basis of the Born approximation.

Our most important result is that the localization length in
semiconductor quantum wires could be many microns long even in modest
quality ({\it i.e.} mean-field $\mu < 10^6cm^2V^{-1}s^{-1}$) samples.
Thus, the physics of Anderson localization is unlikely to adversely
affect operation of microelectronic devices fabricated by using
semiconductor quantum wires.

We acknowledge helpful discussions with Ben Hu. This work is supported
by the U.S.-O.N.R. and the NSF.

\end{document}